# Evaluation of central corneal thickness in keratoconus and normal corneas during air puff indentation

Dan Lin,[a] Lei Tian,[b,*] Chenglang Yuan,[a] Wenxiu Shi,[a] Like Wang,[a] and Yongjin Zhou [a,*]

[a]Shenzhen University, Health Science Center, School of Biomedical Engineering, Xueyuan Avenue 1066, Shenzhen 518055, China

[b]Beijing Institute of Ophthalmology, Beijing Tongren Eye Center, Beijing Tongren Hospital, Capital Medical University;

Beijing Ophthalmology & Visual Sciences Key Laboratory, National Engineering Research Center for Ophthalmology, Beijing 100730, China

*Address all correspondence to: Yongjin Zhou, E-mail: yjzhou@szu.edu.cn; telephone: 0755-26909971; mailing address: Shenzhen University, Health Science Center, School of Biomedical Engineering, Xueyuan Avenue 1066, Shenzhen 518055, China;

Lei Tian, E-mail: tianlei0131@163.comyjzhou@szu.edu.cn; telephone: 0086-18600166885; mailing address: Beijing Ophthalmology & Visual Sciences Key Laboratory, Beijing, 100370, China

*Acknowledgements*

This research was supported by the National Key R&D Program of China (2016YFC0104700), the Science and Technology Planning Project of Guangdong Province (2015A020214022, 2015B020214007), National Natural Science Foundation of China (31600758, 81471735, 61427806), Beijing Natural Science Foundation (7174287), Beijing Nova Program (Z181100006218099), and Beijing Municipal Administration of Hospitals' Youth Programme (QMS20170204).

# Abstract

**Purpose:** This study aimed to investigate the actual changes of central corneal thickness (CCT) in keratoconus and normal corneas during air puff indentation, by using corneal visualization Scheimpflug technology (Corvis ST).

**Methods:** A total of 32 keratoconic eyes and 46 normal eyes were included in this study. Three parameters of $CCT_{initial}$, $CCT_{final}$ and $CCT_{peak}$ were selected to represent the CCT at initial time, final time and highest corneal concavity, respectively, during air puff indentation. Wilcoxon signed rank test (paired sample test) was used to assess the differences between these 3 parameters in both keratoconus and normal groups. Univariate linear regression analysis was performed to determine the effect of $CCT_{initial}$ on $CCT_{peak}$ and $CCT_{final}$, as well as the impact of air puff force on CCT in each group. Receiver operating characteristic (ROC) curves were constructed to evaluate the discriminative ability of the 3 parameters.

**Results:** The results demonstrated that $CCT_{peak}$ and $CCT_{final}$ were significantly decreased ($p<0.01$) compared to $CCT_{initial}$ in both keratoconus and normal groups. Regression analysis indicated a significant positive correlation between $CCT_{peak}$ and $CCT_{initial}$ in normal cornea group ($R^2=0.337$, $p<0.01$), but not in keratoconus group ($R^2=0.029$, $p=0.187$). Likewise, regression models of air puff force and CCT revealed the different patterns of CCT changes between keratoconus and normal cornea groups. Furthermore, ROC curves showed that $CCT_{peak}$ exhibited the greatest AUC (area under ROC curve) of 0.940, with accuracy, sensitivity and specificity of 94.9%, 87.5% and 100%, respectively.

**Conclusions:** CCT may change during air puff indentation, and is significantly different between keratoconus and normal cornea groups. The changing pattern is useful for the diagnosis of keratoconus, and lays the foundation for corneal biomechanics.

## 1 Introduction

Keratoconus is a non-inflammatory corneal disease characterized by progressive thinning of the central cornea [1]. The majority of patients with keratoconus have irregular astigmatism and vision loss [2]. It has been reported that collagen fibrils and interfibers are destroyed by degeneration during the formation of keratoconus [18,20]. These structural changes may alter the biomechanical properties of the cornea [18,19]. Hence, a deep understanding of corneal biomechanics is crucial to describe and even diagnose various types of corneal diseases such as keratoconus [20].

At present, there are only two approaches for the in vivo assessment of corneal biomechanical parameters. Ocular Response Analyzer is the most widely used instrument for measuring the biomechanical properties of human corneas [5,6]. However, it may not be able to display the corneal dynamic deformation process in real time. On the other hand, Corneal Visualization Scheimpflug Technology (Corvis ST) is a relatively newer device that applies a consistent air puff to deform the cornea. The whole process of corneal deformation can be dynamically visualized in real time by using an ultra-high-speed Scheimpflug camera [7,8].

During the deformation process, Corvis ST can measure the changes in corneal biomechanical parameters, such as central corneal thickness (CCT). CCT plays a crucial role in the diagnosis of some corneal diseases, including Fuchs' corneal dystrophy, keratoconus and glaucoma [9-11]. It is believed that the dynamic CCT may contain important information pertaining to various eye diseases associated with endothelial corneal dystrophies or collagen disorders [4]. Unfortunately, the CCT measured by Corvis ST is often fixed and constant, resulting in only few study concerned its dynamic changes during air puff indentation. In addition, the dynamic response of the cornea to an air-puff force remains largely unknown, especially in patients with keratoconus.

Therefore, with the use of corneal dynamic deformation videos from Corvis ST, this study aimed to investigate the actual changing patterns of CCT in keratoconus and normal corneas during air puff indentation. The findings of this study will hopefully guide the clinical use of CCT in diagnosing keratoconus.

**2 Methods**

*2.1 Subject Recruitment*

A comparative cross-sectional study was conducted, involving 32 eyes with keratoconus (keratoconus group) and 46 normal eyes (normal group). For patients with keratoconus in only one eye, the particular eye was selected for measurement. Meanwhile, one eye was randomly selected from normal subjects and patients with keratoconus in both eyes. All patients underwent a complete ophthalmic examination, including a detailed assessment of uncorrected distance visual acuity, corrected distance visual acuity, slit-lamp microscopy and fundus examination, corneal topography (Allegro

Topolyzer; Wavelight Laser Technologie AG, Erlangen, Germany), corneal tomography (Pentacam; Oculus Optikgeräte GmbH), ocular biomechanics, and IOP measurement (Corvis ST). All measurements were taken by two trained ophthalmologists during a single visit. A diagnosis of keratoconus was carried out if the eye had i) an irregular cornea, determined by distorted keratometry mires or distortion of the retinoscopic or ophthalmoscopic red reflex, and ii) at least one of the following slit-lamp signs: Vogt's striae, Fleischer's ring with an arc >2 mm, or corneal scarring consistent with keratoconus [12-14]. Potential subjects who had undergone previous corneal or ocular surgery, had ocular pathology other than keratoconus, and/or had systemic diseases affecting the eyes were excluded from this study.

All participants were asked to remove soft contact lenses for at least 2 weeks and rigid contact lenses for at least 1 month before initial the experiments. Clinical data were collected from March to December 2017 at the Beijing Institute of Ophthalmology, Beijing Tongren Hospital, Beijing, China. This study was approved by the Institutional Review Board of the same hospital. All participants signed a written informed consent form, in accordance with the ethical principles stated in the Declaration of Helsinki.

*2.2 Corvis ST Measurement*

During air puff indentation, the cornea underwent three distinct phases: i) first applanation, ii) the peak concavity and iii) second applanation. A sequence of images of corneal deformation was acquired using Corvis ST (Oculus; Wetzlar, Germany). An ultra-high-speed Scheimpflug camera (4330 frames/s and 8.5 mm horizontal coverage) was used to capture the 139 images of corneal deformation in response to air puff. The final image resolution was 200*576 pixel.

*2.3 Image Processing*

In this study, a robust image processing method was used for the automatic detection of corneal boundaries, as described previously [14]. The flow chart of the corneal boundary extraction process is shown in **Fig. 1**.

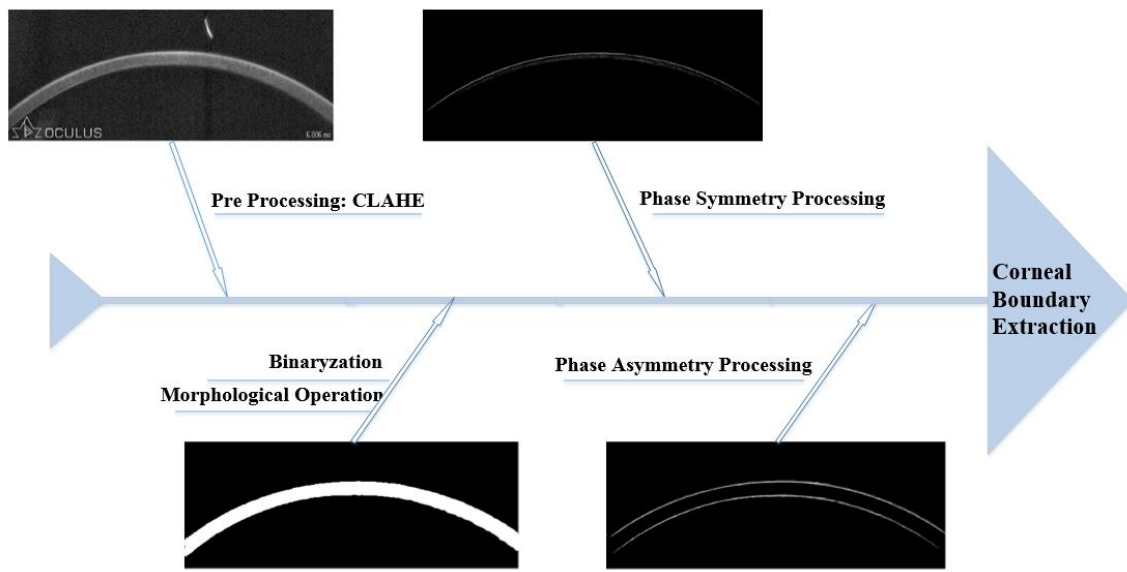

**Fig. 1** Overall flowchart of the proposed method for corneal boundary extraction. First, Contrast Limited Adaptive Histogram Equalization (CLAHE) is used to enhance the contrast of corneal images. Next, binarization, morphological operation and maximum connected region selection are used to extract the corneal contour and remove the artifacts. Finally, phase symmetry processing is used to determine the centerline, while phase asymmetry processing is used to detect the upper and lower corneal boundaries in accordance with the centerline.

As shown in **Fig. 2**, the yellow line indicated the upper boundary and the green line was the lower boundary after phase asymmetry processing. It can be seen that the proposed method was accurate for the extraction of corneal boundaries.

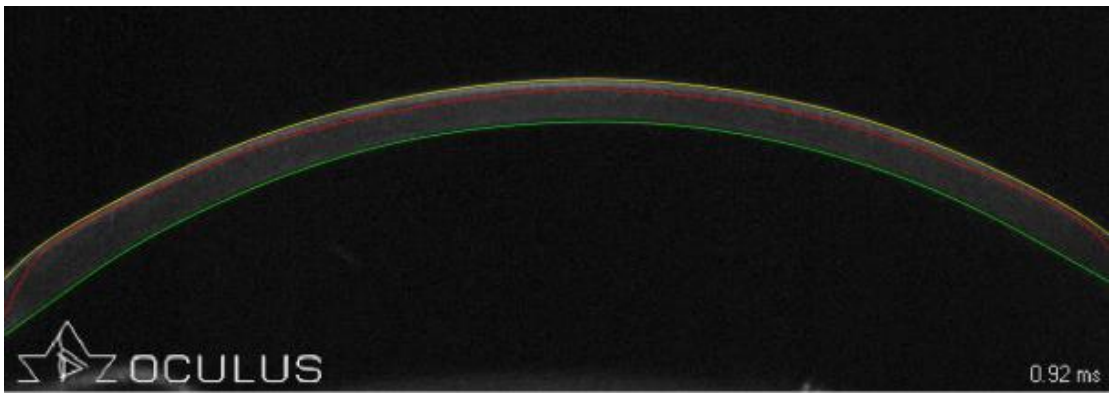

**Fig. 2** Representative figure for the extraction of corneal boundaries. The centerline, upper boundary and lower corneal boundary after fitting are represented by red, yellow and green lines, respectively.

*2.4 CCT Change Measurement*

After extracting corneal boundaries, the corneal thickness at the apex was measured as CCT. A typical curve was constructed for the changes in CCT of keratoconus and normal cornea groups during air puff indentation (**Fig. 3**). Three parameters were used to represent the CCT at different time points during air puff indentation: i) CCT at initial time ($CCT_{initial}$), ii) CCT at peak concavity ($CCT_{peak}$), and iii) CCT at final time ($CCT_{final}$).

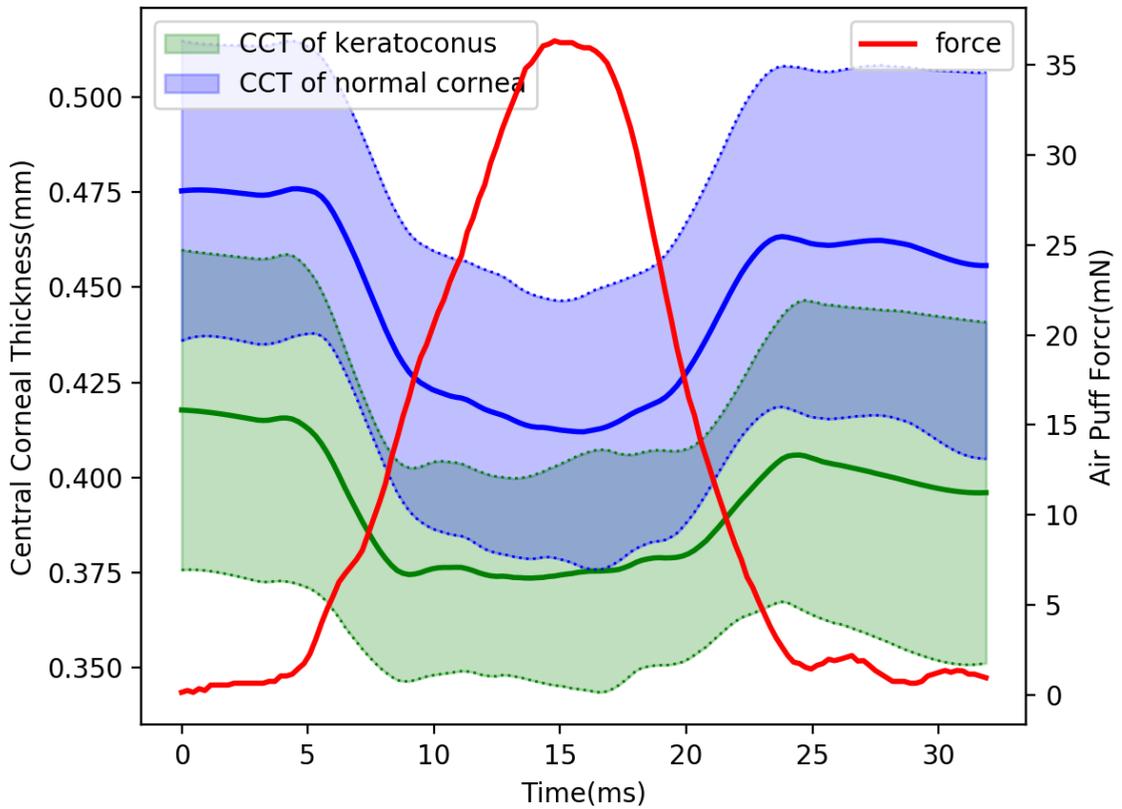

**Fig. 3** A typical representation of CCT time curve during air puff indentation. Blue curves indicate the changes of CCT in normal cornea group, while Green curves represent the changes of CCT in keratoconus group. Standard deviations of CCT in each group are represented by dotted lines. Red curve indicates the force generated by air puff.

*2.5 Statistical Analysis*

All statistical analyses were performed using SPSS (version 19) and Spyder software (Python 3.6). The accuracy of the corneal boundary extraction algorithm was evaluated using Mean Absolute Error (MAE) and Root Mean Square Error (RMSE). The normality assumption of $CCT_{initial}$, $CCT_{peak}$ and $CCT_{final}$ was estimated using

Kolmogorov–Smirnov test. Wilcoxon signed rank test was used to compare the differences between $CCT_{peak}$ and $CCT_{initial}$ as well as $CCT_{final}$ and $CCT_{initial}$ in each group. Mann-Whitney U test was used to compare the differences in these 3 parameters and the proportional limit between keratoconus and normal cornea groups. Univariate linear regression analysis was performed to determine the effect of $CCT_{initial}$ on $CCT_{peak}$ and $CCT_{final}$, and the impact of air puff force on CCT in each group. ROC curve was used to evaluate the diagnostic ability of $CCT_{initial}$, $CCT_{peak}$ and $CCT_{final}$, whereas AUC was used to estimate the predictive ability of these parameters. The fitting degree of linear regression equation was assessed by determination coefficient ($R^2$), while regression coefficient (b) was used to describe how independent variable affects dependent variable. A $p$ values of less than 0.05 were considered statistically significant.

## 3 Results

The results of error analysis demonstrated that the error between the corneal boundaries extracted by manual method and the boundaries extracted by the proposed method was less than 1 pixel (**Table 1**). K-S statistical results showed that $CCT_{initial}$, $CCT_{peak}$ and $CCT_{final}$ did not follow a normal distribution. The results of Wilcoxon signed rank test indicated that the values of $CCT_{peak}$ and $CCT_{final}$ were significantly reduced ($p<0.01$) compared to $CCT_{initial}$ in both keratoconus and normal cornea groups (**Table 2**). The results of Mann-Whitney U test showed that $CCT_{initial}$, $CCT_{peak}$ and $CCT_{final}$ were significantly different ($p<0.01$) between keratoconus and normal cornea groups (**Table 3**).

**Table 1** Error of the corneal boundaries extracted from the proposed method and manual method.

| Error Statistics | Corneal Boundary Extraction method and Manual Method |
|---|---|
| MAE corneal upper border/lower boundary (pixel) | 0.5/1.0 |
| RMSE corneal upper border/lower boundary (pixel) | 0.4/0.5 |

**Table 2** Differences between $CCT_{peak}$ and $CCT_{initial}$ as well as $CCT_{final}$ and $CCT_{initial}$ in both keratoconus and normal cornea groups.

| | Group | Z | $p$ |
|---|---|---|---|
| $CCT_{peak} - CCT_{initial}$ | keratoconus group | -4.6 | <0.001 |
| | normal group | -6.6 | <0.001 |
| $CCT_{final} - CCT_{initial}$ | keratoconus group | -3.7 | <0.001 |
| | normal group | -3.1 | 0.002 |

**Table 3** Comparison of $CCT_{initial}$, $CCT_{peak}$ and $CCT_{final}$ values between keratoconus and normal cornea groups. [mean ± standard deviation (minimum - maximum)]

| Parameters | keratoconus group (n=32) | normal group (n=46) | $p$ |
|---|---|---|---|
| $CCT_{initial}$ (mm) | 0.42±0.04 (0.35~0.52) | 0.49±0.04 (0.39~0.56) | <0.001 |
| $CCT_{peak}$ (mm) | 0.35±0.03 (0.31~0.41) | 0.39±0.03 (0.3~0.46) | <0.001 |
| $CCT_{final}$ (mm) | 0.39±0.04 (0.33~0.51) | 0.47±0.05 (0.37~0.57) | <0.001 |

The relationship between $CCT_{initial}$ and $CCT_{peak}$, $CCT_{final}$ in the two groups were investigated (**Fig. 4**). For normal cornea group, $CCT_{peak}$ and $CCT_{final}$ were positively correlated with $CCT_{initial}$ ($R^2$=0.337; $p$<0.0001 and $R^2$=0.738; $p$<0.0001, respectively). Interestingly, for keratoconus group, $CCT_{initial}$ was strongly positively correlated with $CCT_{final}$ ($R^2$=0.675, $p$<0.0001), but not with $CCT_{peak}$ ($R^2$=0.029, $p$=0.187).

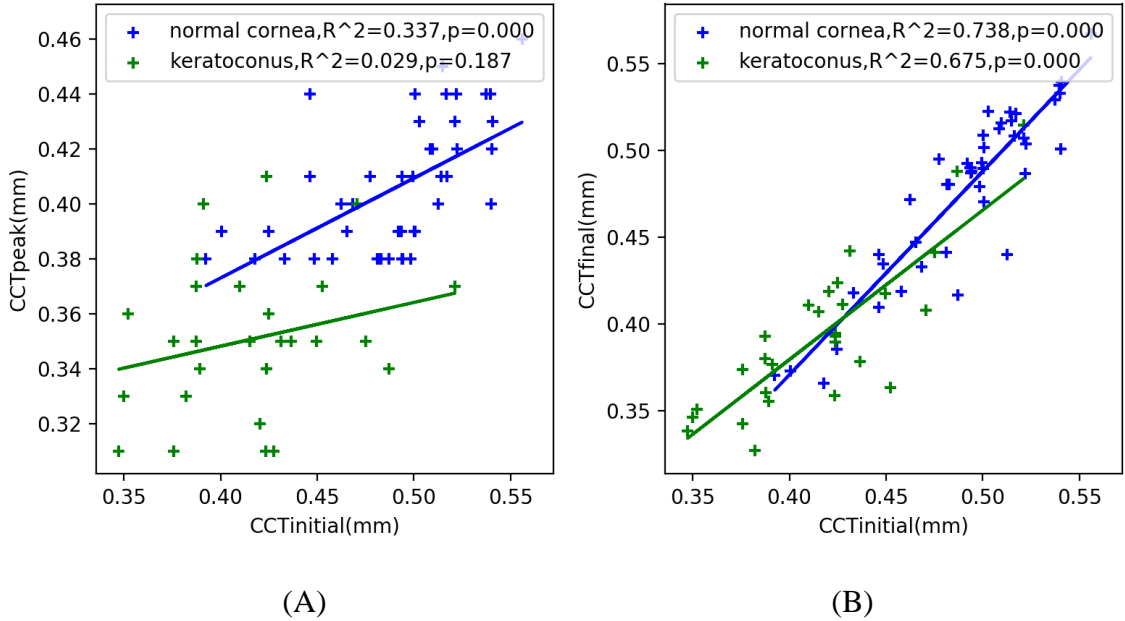

(A)     (B)

**Fig. 4** Linear regressions of $CCT_{initial}$ against $CCT_{peak}$ and $CCT_{final}$ in both keratoconus and normal groups. Blue lines indicate the relationship between these parameters in normal cornea group, while green lines demonstrate the relationship between these parameters in keratoconus group. (A) $CCT_{initial}$ is significantly positively correlated with $CCT_{peak}$ in normal cornea group, but not in keratoconus group. (B) A significant positive correlation is observed between $CCT_{initial}$ and $CCT_{final}$ in both keratoconus and normal cornea groups.

As shown in **Fig. 5**, CCT was almost linearly changed with air puff force in both keratoconus and normal cornea groups. Notably, at higher air puff force, CCT decreased linearly until the force exceeded a specific threshold (proportional limit) and did not return to its original value. The proportional limit of keratoconus group was significantly lower ($p<0.01$) than that of normal cornea group. In addition, CCT decreased more rapidly in keratoconus group than in normal cornea group (b=-0.0030 vs. b=-0.0033, respectively) before reaching its proportional limit.

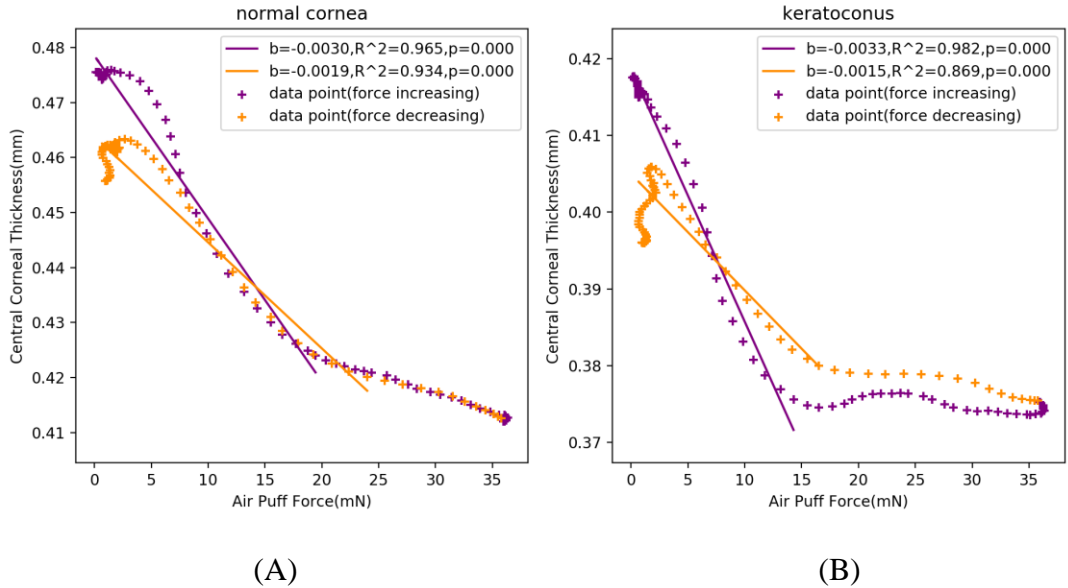

(A)  (B)

**Fig. 5** Linear regressions of air puff force against CCT in both keratoconus and normal cornea groups. Purple curves indicate the changes in CCT during increased air puff force, while the changes of CCT at decreased force are represented by dark orange curves. (A) In normal groups, the proportional limit is approximately 19 mN, and CCT is slowly changed before it. (B) In keratoconus group, the proportional limit is approximately 15 mN, and CCT is rapidly changed before it.

**Table 4** The values of the proportional limit between keratoconus and normal cornea groups. [Mean ± standard deviation (minimum - maximum)]

| Parameters | keratoconus group (n=32) | normal group (n=46) | $p$ |
|---|---|---|---|
| proportional limit (mN) | 18.65±7.34 （8.03~33.22） | 14.84±5.13 （8.03~31.86） | < 0.001 |

ROC curve analysis revealed that $CCT_{peak}$ exhibited the highest AUC of 0.940 compared to $CCT_{initial}$ (AUC=0.900) and $CCT_{final}$ (AUC=0.873), with a cutoff point of

0.5 (**Fig. 6**). The accuracy, sensitivity and specificity of $CCT_{peak}$ were 94.7%, 86.2% and 100%, respectively. These findings suggest that $CCT_{peak}$ is able to discriminate keratoconus from normal corneas.

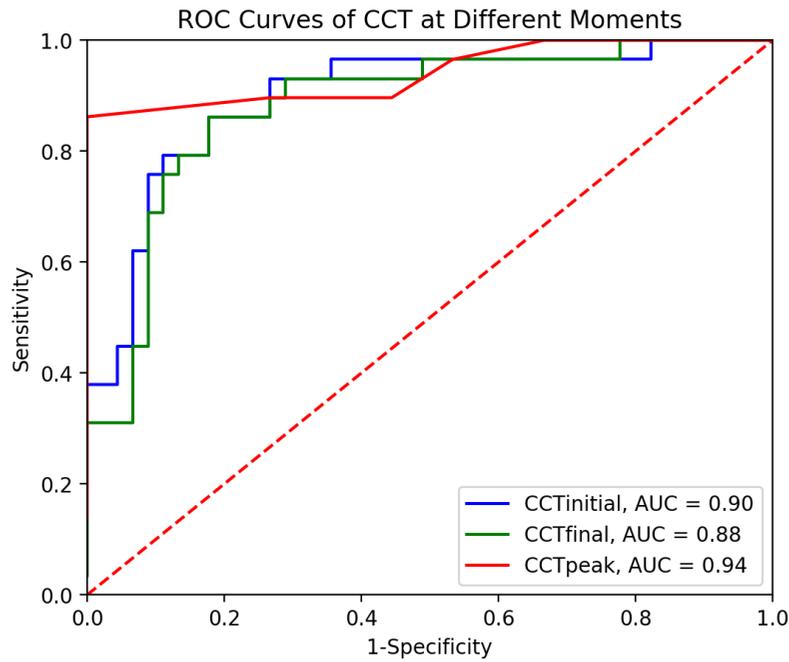

**Fig. 6** ROC curves for $CCT_{initial}$, $CCT_{peak}$ and $CCT_{final}$. $CCT_{peak}$ exhibits the greatest AUC of 0.940, (cutoff point of 0.5), with an accuracy of 94.9%, sensitivity of 87.5% and specificity of 100%.

**Table 5** Classification Table of Keratoconus and Normal Eyes Using $CCT_{initial}$, $CCT_{peak}$ and $CCT_{final}$, Respectively.

| | **$CCT_{peak}$** | | | **$CCT_{initial}$** | | | **$CCT_{final}$** | | |
|---|---|---|---|---|---|---|---|---|---|
| | **Observed** | | | **Observed** | | | **Observed** | | |
| Predicted | Keratoconus | Normal | Overall % | Keratoconus | Normal | Overall % | Keratoconus | Normal | Overall % |
| Keratoconus | 28 | 0 | | 27 | 5 | | 26 | 5 | |

| | | | | | | | | |
|---|---|---|---|---|---|---|---|---|
| Normal | 4 | 46 | | 5 | 41 | | 6 | 41 |
| correct% | 87.5 | 100 | 94.9 | 84.3 | 89.1 | 88.6 | 81.3 | 89.1 | 87.5 |

**4 Discussion**

The alteration of CCT in keratoconus and normal corneas during air puff indentation remains controversial. In Corvis ST software, CCT remains relatively constant for each test. In a previous study [17], CCT at highest concavity has been reduced in healthy corneas. On the contrary, CCT has been found to increase at highest concavity in another study [3]. Therefore, in the present study, an accurate image processing method [14] was used to extract the dynamic CCT during air puff indentation.

As shown in **Table 2**, $CCT_{final}$ and $CCT_{peak}$ were significantly decreased as compared to $CCT_{initial}$ in both keratoconus and normal groups, supporting that cornea may reduce its thickness (compressed) and no longer return to its original level under air puff deformation. These findings reveal that the viscoelastic properties of human cornea are not only manifested by corneal displacement [16], but also the changes in CCT. Additionally, it is worth noting that $CCT_{peak}$ was not significantly associated with $CCT_{initial}$ in keratoconus group (**Fig. 4**). This negative finding reflects the instability in CCT compression, which may be due to the interference of corneal displacement and the reduction of active corneal collagen fibers in patients with keratoconus [18,19].

As aforementioned, the corneal viscoelasticity can be used to reflect the changes in CCT, so it is understandable that CCT may alter its linear decreasing trend as the air-puff force increases. However, the changing pattern of CCT in keratoconus group was different from that of normal cornea group (**Fig. 5**). Thus, it is likely that keratoconus is more readily deformable and less substantial than normal cornea [15,16]. Moreover, it is well documented that keratoconus is thinner than normal cornea [1,15,22]. However, due to the different changing patterns in CCT, the overall thickness difference is possibly enlarged, allowing the maximum AUC for $CCT_{peak}$ instead of $CCT_{initial}$. The exact of those differences is still unclear and needs further research. If this observation is proven to be true, $CCT_{peak}$ should be taken into consideration in the clinical practice of keratoconus diagnosis, through the combination of multiple corneal biomechanical parameters, including $CCT_{initial}$.

Interestingly, both compression and displacement can occur simultaneously in the vitreous body under the pressure of injected perfluoropropane gas [20]. However, in the case of Corvis ST air puff indentation, only displacement is expected to take place, rather than both of them. Besides, compressive viscoelasticity has been long recognized as a crucial biomechanical index of cornea [21]. The reason for omitting cornea compression is probably due to the relatively small changes in CCT (**Fig. 3**) restricted by the low imaging resolution of Corvis ST. Though beyond the scope of this paper, the enhancement of imaging resolution as well as the precise measurement precision of CCT, in future, may be important for the noninvasive assessment of corneal biomechanical properties.

There are some unavoidable limitations of this study, including the limited accuracy of corneal boundary extraction method (**Table 1**). Indeed, the proposed method [14] had a minor error as compared to the manual method, which may eventually affect the accuracy of dynamic CCT measurement. Moreover, the sample size in this study was relatively small, and therefore all the statistical results must be interpreted cautiously.

**5 Conclusion**

In sum, CCT is changed to adapt the stress generated by Corvis air puff, and thus can be used to identify keratoconus. Therefore, this study is important for the fields of corneal biomechanics and the diagnosis of different corneal diseases such as keratoconus. Further studies with a more precise method of corneal boundary extraction and larger collection of datasets are warranted, for the purpose of developing and implementing CCT.

*Disclosures*

The authors have no relevant financial interests in this study and no potential conflicts of interest to disclose.

**Fig. 1** Overall flowchart of the proposed method for corneal boundary extraction.

**Fig. 2** Representative figure for the extraction of corneal boundaries.

**Fig. 3** A typical representation of CCT time curve during air puff indentation.

**Fig. 4** Linear regressions of $CCT_{initial}$ against $CCT_{peak}$ and $CCT_{final}$ in both keratoconus and normal groups.

**Fig. 5** Linear regressions of air puff force against CCT in both keratoconus and normal cornea groups.

**Fig. 6** ROC curves for $CCT_{initial}$, $CCT_{peak}$ and $CCT_{final}$. $CCT_{peak}$ exhibits the greatest AUC of 0.940, (cutoff point of 0.5), with an accuracy of 94.7%, sensitivity of 86.2% and specificity of 100%.

**Table 1** Error of the corneal boundaries extracted from the proposed method and manual method.

**Table 2** Differences between $CCT_{peak}$ and $CCT_{initial}$ as well as $CCT_{final}$ and $CCT_{initial}$ in both keratoconus and normal cornea groups.

**Table 3** Comparison of $CCT_{initial}$, $CCT_{peak}$ and $CCT_{final}$ values between keratoconus and normal cornea

groups.

**Table 4** The values of the proportional limit between keratoconus and normal cornea groups.

**Table 5** Classification Table of Keratoconus and Normal Eyes Using $CCT_{initial}$, $CCT_{peak}$ and $CCT_{final}$, Respectively.